\documentclass[epjCONF]{svjour}
\usepackage{graphicx}
\usepackage[varg]{txfonts} 
\usepackage[latin1]{inputenc}
\session-title{Conference Title, to be filled}
\begin{document}
\title{Tidal Streams in Newly Discovered compact elliptical (cE) galaxies}
\author{A. P. Huxor \inst{}\fnmsep\thanks{\email{avon.huxor@bristol.ac.uk}} \and  S. Phillipps \inst{} \and J. Price \and  R. Harniman}
\institute{University of Bristol, H. H. Wills Physics Laboratory, Tyndall Avenue, Bristol, BS8 1TL, U.K.}
\abstract{
We present two newly-discovered, compact elliptical (cE) galaxies, which exhibit clear evidence of tidal steams, found in a search of SDSS DR7. The structural parameters of the cEs are derived using GALFIT and give effective radii $<$ 400 pc. They also possess  young to intermediate-age stellar populations. These two cEs provide direct evidence,   a "smoking gun'',  for the process of tidal stripping  that is believed to be the origin of M32-type galaxies. Both are found in small groups, suggesting that we may be seeing the formation of such galaxies in dynamically young galaxy groupings.} 
\maketitle
\section{Introduction}
\label{intro}
There has been much debate over the origins of the so-called compact ellipticals.  They are therefore generally suspected to be the result of tidal stripping and truncation driven by interactions with their giant neighbours (\cite{Faber73,Bekkietal01}). An alternative view suggests that they are the low-luminosity end of the family of normal elliptical galaxies, in which the high surface brightness is the result of earlier star formation triggered by dissipative  mergers (\cite{Kormendyetal09}).  
The evidence for stripping is still somewhat indirect, comprising faint hints of tidal features (e.g. \cite{SmithCastellietal08}), and data showing that cEs having colours and velocity dispersions consistent with more massive galaxies, and thus supporting a scenario in which the latter are stripped leaving cE remnants (e.g. \cite{Priceetal09}). To try to study the possible formation channels of cEs , we have searched for cE candidates in SDSS DR7, up to a redshift of 0.025, and have a sample of ~ 60 candidate objects (Huxor et al., in prep), many of which are undergoing follow-up. A handful of objects, however, are notable in themselves.

Among our candidates we have two cEs which show clear evidence of tidal stripping in action $-$ the smoking gun of cE formation by this channel. In one case (cE2) we have Megacam archival data. Both cE1 and cE2 present clear tidal tails, and are very marked in cE2, in which the major tail contains a significant proportion  of the amount of flux found in the cE itself. The morphology of these tails points to the progenitors being spiral galaxies, the tails being the remnant of the dynamically cold disk.
The structural properties of these galaxies were obtained with GALFIT. All our cEs have SDSS spectra, and PPXF was employed to obtain velocity dispersions, followed by EZ-Ages to derive ages and metallicities.

\begin{figure}
\resizebox{0.75\columnwidth}{!}
{%
 \includegraphics{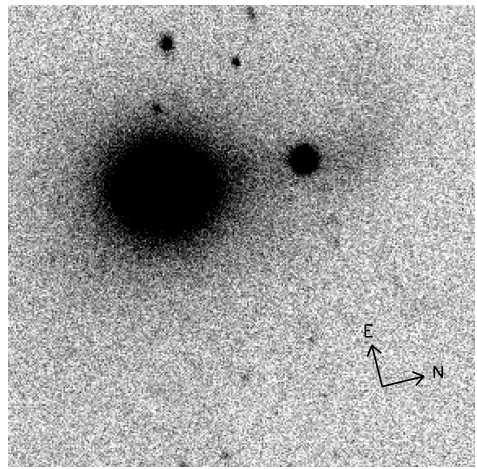} }
\caption{SDSS r-band Image of cE1, scaled to enhance the tidal tails. 50 kpc  $\times$ 50 kpc. The cE is not centred in the image, as it lies on the edge of the CCD.}
\label{Fi:cE1}       
\end{figure}

\begin{figure}
\resizebox{0.75\columnwidth}{!}{%
 \includegraphics{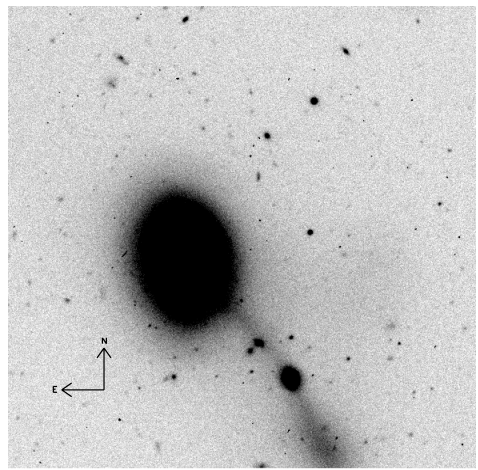} }
\caption{CFHT/Megacam r-band image of cE2 scaled to enhance the tidal tails. 50 $\times$ 50 kpc. It too lies at the edge of the CCD.}
\label{Fi:cE2}       
\end{figure}
\begin{figure*}
 \resizebox{1.6\columnwidth}{!}{
 \includegraphics[angle=90]{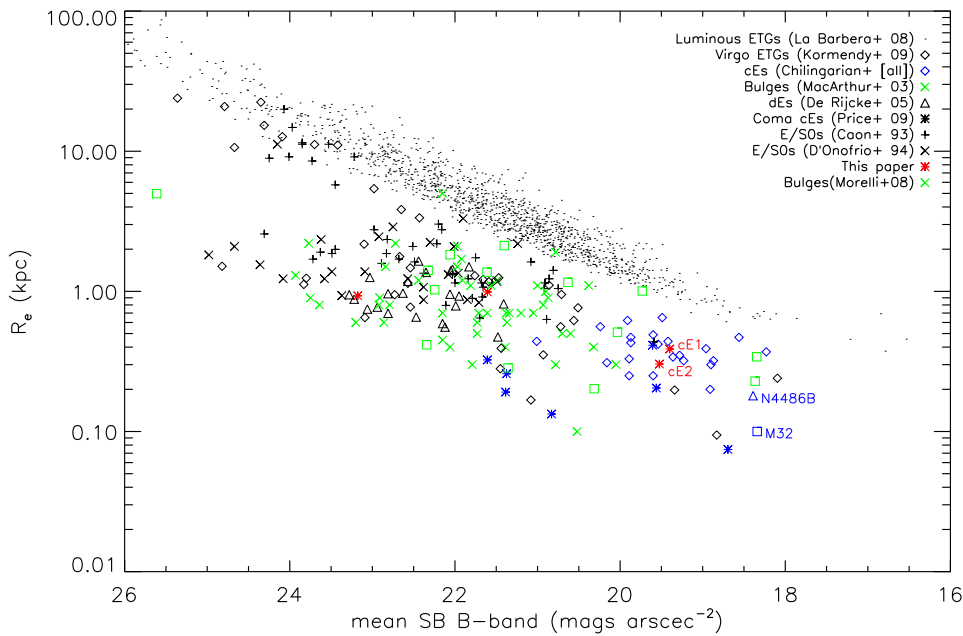} }
\caption{ Plot of the mean surface brightness within $R_{e}$  (effective radius) against  $R_{e}$ for the new cEs (using the SDSS discovery data), compared with other spheroidal stellar systems. The new cEs are labelled. Also shown are the two other examples (unlabelled red symbols) known of (more massive) dwarfs in the process of being stripped (\cite{Forbesetal03,Sasakietal07}) .}
\label{Fi:plot}       
\end{figure*}

\section{Results}
\label{results}

The results are illustrated in Figure \ref{Fi:plot}. The two new galaxies clearly lie in in the region of the plot occupied by the cEs. However, it is notable that they also share this region of the plot with bulges.  Both cEs reveal an SSP eqivalent age that is relatively young,  $\sim$ 3.6 -- 5.4 Gyr for cE1 and cE2 respectively.
The presence of a younger stellar population makes these objects somewhat more akin to the the prototype cE  (M32) than other cE in the literature; for example,  \cite{Schiavonetal04} give a spectroscopic age for M32 of between 2.0 and 3.5 Gyr . However, \cite{Chilingarianetal09} note that none of their cEs exhibits a young stellar population.

\section{Discussion}
\label{discussion}

One interesting result is the environments in which cE1 and cE2 are found. Conventionally, searches for cEs have focussed on galaxy clusters (e.g. \cite{Chilingarianetal09}), on the premise that the  cluster environment is more favourable to stripping. However, cE1 and cE2 are in  small groups. This can be partly explained by the more rapid dissolution of streams in the complex tidal fields of a cluster (\cite{Rudicketal09}), and by the presence of intra-cluster light making detection of tidal features difficult. In addition, there are also more galaxies in groups than in clusters (\cite{Ekeetal05}).  However, we may be seeing a significant feature.
Both cEs are  in groups that have a high fraction of late-type galaxies, suggesting that these groups are in a young dynamic state,. We may be seeing an example of pre-processing, in which galaxies are transformed in a group environment prior to their possible later assembly into a cluster. It is possible that many of the cEs found in galaxy clusters may have formed this manner and this may explain the old ages for some of the cEs reported by  \cite{Chilingarianetal09}.


\begin{thebibliography}{}




\bibitem{Faber73} Faber S.~M., 1973, ApJ, 179, 423

\bibitem{Bekkietal01} Bekki K., Couch W.~J., Drinkwater M.~J., Gregg M.~D., 2001, ApJ, 557, L39


\bibitem{Kormendyetal09} Kormendy J., Fisher D.~B., Cornell M.~E., Bender R., 2009, ApJS, 182, 216 

\bibitem{SmithCastellietal08} Smith Castelli A.~V., Faifer F.~R., Richtler T., Bassino L.~P., 2008, MNRAS, 391, 685 

\bibitem{Priceetal09} Price J., et al., 2009, MNRAS, 397, 1816

\bibitem{Forbesetal03} Forbes D.~A., Beasley M.~A., Bekki K., Brodie J.~P., Strader J., 2003, Sci, 301, 1217

\bibitem{Sasakietal07} Sasaki S.~S., et al., 2007, ApJS, 172, 511

\bibitem{Schiavonetal04} Schiavon R.~P., Caldwell N., Rose J.~A., 2004, AJ, 127, 1513

\bibitem{Chilingarianetal09} Chilingarian I., Cayatte V., Revaz Y., 
Dodonov S., Durand D., Durret F., Micol A., Slezak E., 2009, Sci, 326, 1379 

\bibitem{Rudicketal09} Rudick C.~S., Mihos J.~C., Frey L.~H., McBride C.~K., 2009, ApJ, 699, 1518 

\bibitem{Ekeetal05} 
Eke V.~R., Baugh C.~M., Cole S., Frenk C.~S., King H.~M., Peacock J.~A., 
2005, MNRAS, 362, 1233


\end{thebibliography}
\end{document}